# Fracture properties of La(Fe,Mn,Si)$_{13}$ magnetocaloric materials

Siyang Wang[a,*], Paul Burdett[b], Edmund Lovell[b], Rachel Bettles[b], Neil Wilson[b], Mary P. Ryan[a], Finn Giuliani[a]

[a]Department of Materials, Imperial College London, London, SW7 2AZ, UK

[b]Camfridge Ltd., Cambridge, CB22 3GN, UK

*Email: siyang.wang15@imperial.ac.uk

## Abstract

La(Fe,Mn,Si)$_{13}$ alloys are a promising material family for magnetic refrigeration. Challenges associated with their structural integrity during device assembly and operation requires deep understanding of the mechanical properties. Here we developed a workflow to quantitatively study the fracture properties of La(Fe,Mn,Si)$_{13}$ plates used in magnetic cooling devices. We employed microstructural characterisation, optical examination of defects, and four-point bending tests of samples with known defect sizes to evaluate their mechanical performance. We established the residual strength curve which directly links observed defects to mechanical strength. The estimated fracture toughness $K_C$ of hydrogenated La(Fe,Mn,Si)$_{13}$ is approximately 4 MPa·m$^{1/2}$ for the geometry employed. The established relationship between strength and crack length enables the prediction of mechanical performance through examination of defects *via* optical microscopy, therefore can be used industrially for directing plate selection to guarantee the mechanical stability of refrigeration devices.

Magnetic refrigeration could drastically reduce $CO_2$ emission by the cooling sector through reduced electricity consumption, due to the potential high energy efficiency of this technology [1,2]. A key challenge in its commercialisation arises from the poor mechanical stability of the La(Fe,Mn,Si)$_{13}$ magnetocaloric materials when subjected to simultaneous action of several generalised forces during processing and service, leading to limited lifetime [3,4]. Understanding the mechanical properties of these materials is therefore of high demand, yet studies are limited [5,6]. We found previously that although the intrinsic strength of La(Fe,Mn,Si)$_{13}$ is up to ~6 GPa, strength of the actual materials in use is reduced by a factor of ~20 likely due to cracks/pores in the microstructure [6,7] and potentially (pre-existing) dislocations and grain boundaries. While complete removal of the cracks is difficult because of technical limitations and service environment requirements, it is important to understand the tolerance of cracks in the material for prolonged stable performance, thereby establishing a guideline for quality control before device production.

The strength of La(Fe,Si)$_{13}$ during compression varies with sample dimension [5], in line with other defect-controlled materials. This highlights the importance of carrying out mechanical testing on samples with dimensions identical to those employed in cooling devices, thereby providing relevant properties that can be used to accurately predict in-service performance. This is however difficult, as materials in use are often small and thin wafers (akin to human nails in dimensions) with high surface area to volume ratio for high energy efficiency, making them impossible to fit onto conventional mechanical testing platforms. Bespoke testing fixture is thus necessary, and bending is a realistic geometry for testing such plate specimens. Three-point bending is not ideal for exploring the effect of cracks on mechanical properties, as the strength extracted is only sensitive to cracks within a small volume at the middle of the specimens due to concentrated strain energy. Four-point bending is advantageous as it subjects a wider area on the specimen to a more uniform stress field. We therefore employ four-point bending for this study, to understand the variation of strength with (pre-existing) crack length, and to estimate toughness.

La(Fe,Mn,Si)$_{13}$ plates (hydrogenated, with the temperature where the magnetocaloric effect is maximum, $T_{peak}$ – as measured by adiabatic temperature change for 0 to 1.5 T field change – of 20.9 °C) with identical dimensions to those employed in magnetic cooling devices were received from Camfridge Ltd. The plate thickness is in the sub-mm scale, and the thickness to width ratio is ~1:30. Microstructure of the plates was characterised using scanning electron microscopy (SEM) imaging and electron backscatter diffraction (EBSD), on a FEI Quanta 650 SEM with a Bruker eFlashHR (v2) EBSD camera using a beam voltage of 20 kV [6]. As shown in Figure 1, the material consists of two main phases: the La(Fe,Mn,Si)$_{13}$ (also termed main/1:13) phase and the α-Fe phase. The nominal volume fraction of the α-Fe phase is 15%. The pole figures indicate an absence of texture in the main phase. The composition of the 1:13 phase, as measured by SEM-energy dispersive X-ray spectroscopy (SEM-EDX), is LaFe$_{10.7}$Mn$_{0.4}$Si$_{1.2}$.

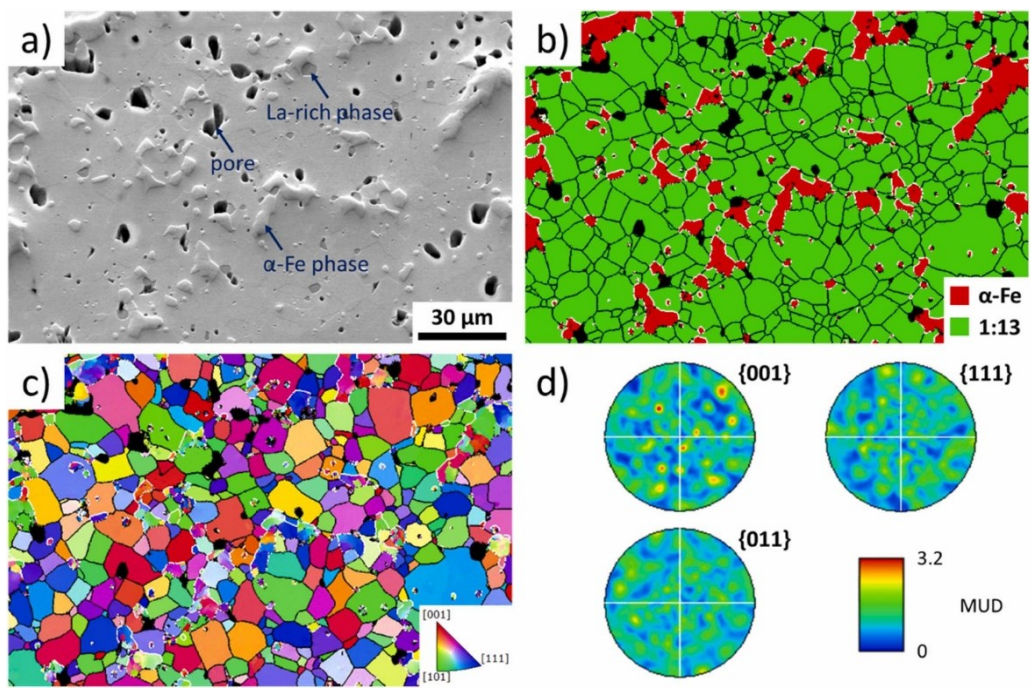

Figure 1 (a) Secondary electron image, (b) phase map, (c) inverse pole figure (IPF)-out-of-page map, and (d) pole figures (for the 1:13 phase) of an area on the as-received specimen. In (b) and (c), grain boundaries of the 1:13 phase, and phase boundaries between La(Fe,Mn,Si)$_{13}$ and α-Fe are highlighted with black and white lines, respectively. MUD denotes multiples of the uniform density [6]. Note that the image in (a) shows the cross sections of the pores, of which the 3D shapes could be revealed by X-ray or focussed ion beam tomography.

A bespoke four-point bending fixture was designed and fabricated (Figure 2(a)), which enabled testing of the plates on a Gatan/Deben Microtest 300 mechanical testing stage. The loading span was designed to be ½ of the support span. Before mechanical testing, optical microscopy examination of each individual plate (both sides) was conducted, in order to pick out plates with (typically) one pre-existing through-thickness crack present within the region that will be subjected to the stress field between the two loading/inner pins during the tests. The cracks were likely generated upon cutting and/or handling of the plates, and their growth may be retarded by the α-Fe phase [4]. The length of each crack was measured. Five plates without observed pre-existing cracks were tested for comparison. Tests were performed with a displacement speed of 0.1 mm/min. Figure 2(b) shows a typical load-displacement response of the plates during the tests, where fracture occurred immediately after linear elastic loading without plastic flow, as per prior three-point bending tests [6].

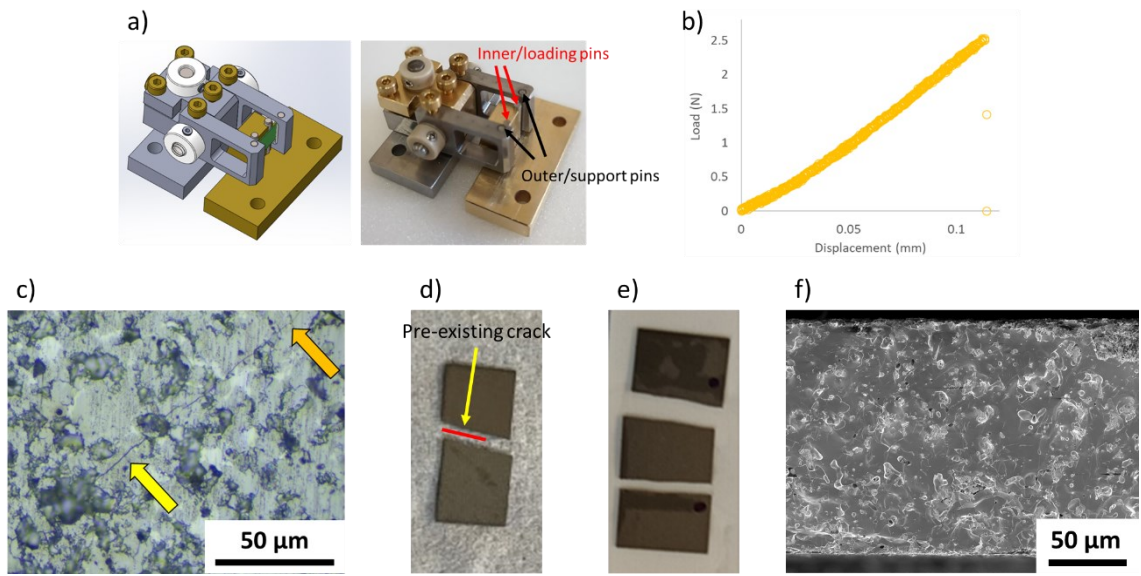

Figure 2 (a) The bespoke four-point bending fixture for testing the plate specimens. (b) A typical load-displacement curve for the plates recorded during the tests. (c) Optical micrograph showing a pre-existing crack in a plate, and pictures of (d) a plate with a pre-existing crack and (e) a plate without observable pre-existing cracks after four-point bending tests. (f) SEM image of the fracture surface for a sample post-test.

Figure 2(c) shows an optical micrograph of a plate captured before the tests revealing a pre-existing crack. After the tests, plates with pre-existing cracks all fracture into two pieces. The cracks were found to develop along the pre-existing cracks, and an example is shown in Figure 2(d). In contrast, plates without observable pre-existing cracks broke into three nearly even pieces, where the two cracks were both within the region between the loading/inner pins for all the plates (Figure 2(e)). The fracture mode is mainly intragranular as evidenced by Figure 2(f), in agreement with our prior work on hydrogenated La(Fe,Mn,Si)$_{13}$ [6]. We found previously that hydrogenation changes the fracture mode of this material upon bending tests [6], but we note that in-service fracture mode may also be affected by factors such as corrosive media [7] and/or external stress state [8].

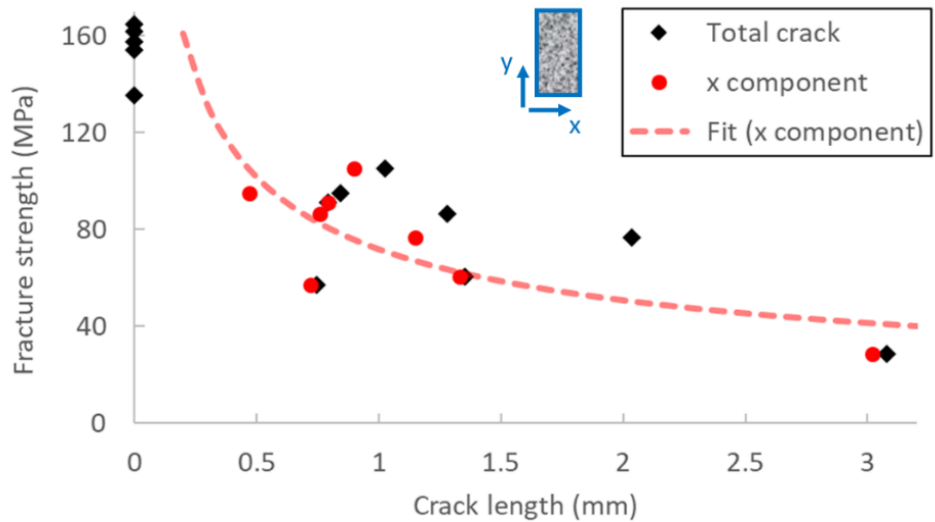

Figure 3 Variation of fracture strength of the plates as a function of total crack length (black) and *x* component of crack length (red) which is the component of crack length along the direction perpendicular to principal stress direction during four-point bending tests (*x* and *y* axes are illustrated in the insert). The dashed line is the fit of the data for the fracture strength vs. *x* component of the crack length to Equation 1.

The variation of plate strength with pre-existing crack length, or residual strength curve, is shown in Figure 3. Plates without observable pre-existing cracks exhibit fracture strengths of 130-170 MPa, in agreement with prior work [6]. Broadly, as the crack length increases, the fracture strength decreases. This graph serves as a guideline for quality control of plates used to assemble regenerators. The established relationship between strength and crack length enables the prediction of mechanical performance through examination of defects *via* optical microscopy, therefore can be used for directing plate selection to guarantee the mechanical stability of refrigeration devices. Note that residual strength curves for brittle materials are often accessed *via* analytical prediction based on the theory of fracture mechanics and material properties, for example [9]. Instead, rarely have they been established directly through experimental measurements to our best knowledge, and only few data sources such as [10] are available. This likely arises from the difficulty in obtaining well-defined samples with a broad distribution of defect sizes. However, we have shown here that through careful examination and testing of the plates, direct establishment of the residual strength curve can be achieved for this particular material and geometry, and is vastly useful for controlling in-service performance.

If strength is controlled by brittle fracture under (far-field) tensile stress, then the variation of strength with crack size should generally follow Equation 1 [11].

$$\sigma = \frac{K_C}{\sqrt{\pi a}} \qquad 1$$

where $\sigma$ is the fracture strength, $K_C$ the fracture toughness, and $a$ the crack length.

Fitting the data for the fracture strength vs. $x$ component of the crack length to Equation 1 (dashed line in Figure 3) gives an estimated $K_C$ of 4 MPa·m$^{1/2}$, which falls into the characteristic range of brittle materials [12]. Note that this is only a rough estimation of $K_C$, given the stress state being not uniform tensile and the crack length being not negligible compared to plate width. However, this value extracted aligns with the load-displacement behaviour (Figure 2(b)) which indicates the brittle nature of the material, and prior tests where even µm-scale test pieces failed in a quasi-brittle manner [6].

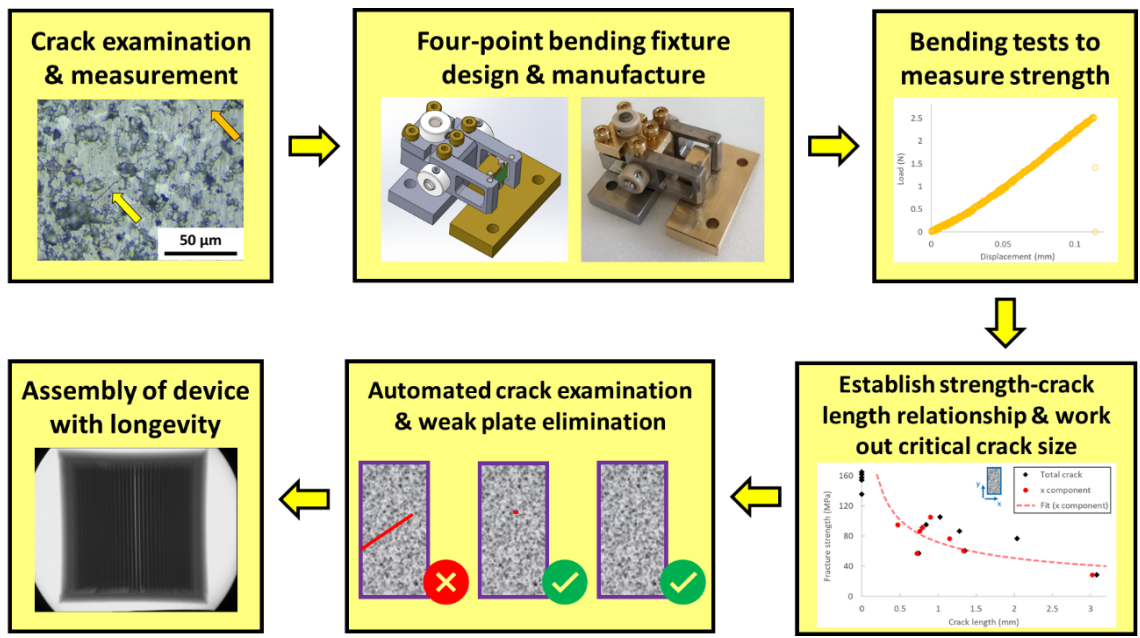

Figure 4 Summary of the workflow established in this work to quantify the correlation between fracture strength and pre-existing crack length, which can be used for quality control and mechanical stability prediction of regenerators in magnetic refrigeration systems.

In summary, we established a workflow (Figure 4) which enabled quantification of the correlation between fracture strength and pre-existing crack length of La(Fe,Mn,Si)$_{13}$ magnetocaloric material. The results, obtained through testing plate specimens with identical geometry to those used in magnetic cooling devices, can be directly used for quality control and mechanical stability prediction of regenerators in magnetic refrigeration systems. For the material studied, evaluation of the fracture toughness was achieved for the first time. An estimated $K_C$ of 4 MPa·m$^{1/2}$ for the testing geometry employed indicates the brittle nature of the material.

## Acknowledgements

We acknowledge funding from Innovate UK (UKRI 32645).

## References

[1] M. Balli, S. Jandl, P. Fournier, A. Kedous-Lebouc, Advanced materials for magnetic cooling: Fundamentals and practical aspects, Appl. Phys. Rev. 4 (2017). https://doi.org/10.1063/1.4983612.

[2] V. Franco, J.S. Blázquez, J.J. Ipus, J.Y. Law, L.M. Moreno-Ramírez, A. Conde, Magnetocaloric effect: From materials research to refrigeration devices, Prog. Mater. Sci. 93 (2018) 112–232. https://doi.org/10.1016/j.pmatsci.2017.10.005.

[3] S. Lionte, A. Barcza, M. Risser, C. Muller, M. Katter, LaFeSi-based magnetocaloric material analysis: Cyclic endurance and thermal performance results, Int. J. Refrig. 124 (2021) 43–51. https://doi.org/10.1016/j.ijrefrig.2020.12.004.

[4] S. Wang, J.O. Douglas, E. Lovell, N. Wilson, L. Guo, B. Gault, M.P. Ryan, F. Giuliani, Near-atomic scale chemical analysis of interfaces in a La(Fe,Mn,Si)13-based magnetocaloric material, Scr. Mater. 224 (2023) 115143. https://doi.org/10.1016/j.scriptamat.2022.115143.

[5] O. Glushko, A. Funk, V. Maier-Kiener, P. Kraker, M. Krautz, J. Eckert, A. Waske, Mechanical properties of the magnetocaloric intermetallic LaFe11.2Si1.8 alloy at different length scales, Acta Mater. 165 (2019) 40–50. https://doi.org/10.1016/j.actamat.2018.11.038.

[6] S. Wang, O. Gavalda-Diaz, T. Luo, L. Guo, E. Lovell, N. Wilson, B. Gault, M.P. Ryan, F. Giuliani, The effect of hydrogen on the multiscale mechanical behaviour of a La(Fe,Mn,Si)13-based magnetocaloric material, J. Alloys Compd. 906 (2022) 164274.

https://doi.org/10.1016/j.jallcom.2022.164274.

[7] S. Pan, J. Yuan, C. Linsley, J. Liu, X. Li, Corrosion behavior of nano-treated AA7075 alloy with TiC and TiB2 nanoparticles, Corros. Sci. 206 (2022) 110479. https://doi.org/10.1016/j.corsci.2022.110479.

[8] X. Yang, S. Gao, Analysis of the crack propagation mechanism of multiple scratched glass-ceramics by an interference stress field prediction model and experiment, Ceram. Int. 48 (2022) 2449–2458. https://doi.org/10.1016/j.ceramint.2021.10.026.

[9] Z.H. Jin, R.C. Batra, Some basic fracture mechanics concepts in functionally graded materials, J. Mech. Phys. Solids. 44 (1996) 1221–1235. https://doi.org/10.1016/0022-5096(96)00041-5.

[10] T.W. Orange, Fracture Toughness of Wide 2014-T6 Aluminum Sheet at -320 °F, Washington, D. C., 1967. https://ntrs.nasa.gov/search.jsp?R=19670018968.

[11] N.E. Dowling, K.S. Prasad, R. Narayanasamy, Mechanical Behavior of Materials : Engineering Methods for Deformation, Fracture, and Fatigue, 4th & intl ed., Pearson, Boston, Mass., 2013.

[12] G.A. Gogotsi, Fracture toughness of ceramics and ceramic composites, Ceram. Int. 29 (2003) 777–784. https://doi.org/10.1016/S0272-8842(02)00230-4.